\documentclass{article}
\usepackage{microtype}

\usepackage{xspace}
\usepackage{amsmath}
\usepackage{amsthm}
\usepackage{amssymb}
\usepackage[colorinlistoftodos]{todonotes}
\usepackage[nocompress]{cite}
\usepackage{authblk}

\newcommand{\sseh}{\textsc{Small-Set-Expansion Hypothesis}\xspace}
\newcommand{\sse}{\textsc{Small-Set-Expansion}\xspace}
\newcommand{\ssehard} {\textsc{Small-Set-Expansion-hard}\xspace}
\newtheorem{theorem}{Theorem}[section]
\newtheorem{lemma}[theorem]{Lemma}
\newtheorem{corollary}[theorem]{Corollary}
\newtheorem{conj}[theorem]{Conjecture}
\newtheorem{definition}[theorem]{Definition}

\newcommand{\E}{\mbox{\bf E}}

\newcommand{\eps}{{\epsilon}}

\newcommand*{\R}{\mathbb{R}}
\newcommand*{\trans}{^{\mathrm{T}}}
\newcommand*{\xs}{{x_S}}
\newcommand*{\N}{\mathbb{N}}

\title{Computational Complexity of Certifying Restricted Isometry Property}

\author[1]{Abhiram Natarajan\thanks{nataraj2@purdue.edu}}
\author[1]{Yi Wu\thanks{wuyi@purdue.edu}}
\affil[1]{Department of Computer Science, Purdue University, USA}

\begin{document}

\maketitle
\begin{abstract}
Given a matrix $A$ with $n$ rows, a number $k<n$, and $0<\delta < 1$, $A$ is $(k,\delta)$-RIP (Restricted Isometry Property) if, for any vector $x\in \R^n$, with at most $k$ non-zero co-ordinates, $$(1-\delta) \|x\|_2 \leq \|A x\|_2 \leq (1+\delta)\|x\|_2$$

In other words, a matrix $A$ is $(k,\delta)$-RIP if $A x$ preserves the length of $x$ when $x$ is a $k$-sparse vector. In many applications, such as compressed sensing and sparse recovery, it is desirable to construct RIP matrices with a large $k$ and a small $\delta$. It is known that, with high probability, random constructions produce matrices that exhibit RIP. This motivates the problem of certifying whether a randomly generated matrix exhibits RIP with suitable parameters.

In this paper, we prove that it is hard to approximate the RIP parameters of a matrix assuming the \sseh. Specifically, we prove that for any arbitrarily large constant $C>0$ and any arbitrarily small constant $0<\delta<1$, there exists some $k$ such that given a matrix $M$, it is \ssehard to distinguish the following two cases:

\begin{itemize}
\item (Highly RIP) $M$ is $(k,\delta)$-RIP.
\item (Far away from RIP) $M$ is not $(k/C, 1-\delta)$-RIP.
\end{itemize}

Most of the previous results on the topic of hardness of RIP certification only hold for certification when $\delta=o(1)$; i.e, when the matrix exhibits strong RIP. In practice, it is of interest to understand the complexity of certifying a matrix with $\delta$ being close to $\sqrt{2}-1$, as it suffices for many real applications to have matrices with $\delta = \sqrt{2}-1$. Our hardness result holds for any constant $\delta$. Specifically, our result proves that even if $\delta$ is indeed very small, i.e. the matrix is in fact \emph{strongly RIP}, certifying that the matrix exhibits \emph{weak RIP} itself is \ssehard.

In order to prove the hardness result, we prove a variant of the Cheeger's Inequality for sparse vectors. Although a similar result is already known, our proof technique gives better constants in the inequality which may be useful for other applications. Specifically, let $A$ be the adjacency matrix of a $d$-regular graph $G(V,E)$, and $L = I- \frac{1}{d}A$ be the normalized Laplacian matrix of $G$. For any $\eta \leq 1/2$, we show that $${\lambda_{\eta}} \le \phi_{\eta}(G) \leq   \sqrt{(2-\lambda_\eta)\lambda_\eta}$$ where $\lambda_\eta= \min_{\|x\|_0=\eta n} \frac{x^T  L x}{|x|^2_2}$ and $\phi_{\eta}(G)$ is the minimum edge expansion among all the sets of size at most $\eta |V|$.

It is interesting to note that the relationship between $\lambda_{\eta}$ and $\Phi_{\eta}(G)$ is different from (and tighter than) the relation between $\lambda$ and $\phi(G)$ in the regular version of Cheeger's Inequality (which states that $\frac{\lambda}{2} \leq \phi(G)\leq \sqrt{2\lambda}$). We will see that obtaining this tighter relationship between $\lambda_\eta$ and $\phi_\eta(G)$ is crucial in proving our hardness result.
\end{abstract}

\section{Introduction}
Moore's law has enabled the creation of very robust and effective sensing systems. As a result of the ubiquity of such systems, the amount of data generated by these systems has increased vastly. In fact, in most real applications, there is so much data that sampling at the required rates (called Nyquist rate) becomes impractical due to data storage problems as well as the sheer magnitude of the sampling rate~\cite{davenport2011introduction}. Signal processing literature shows us that this problem is circumvented by constructing compressible representations of signals. This technique leverages the notion of \emph{sparse approximation} and is called \emph{compressed sensing}.

A formal statement of the central problem of compressed sensing is as follows. Assume the presence of a matrix $\Phi \in \R^{m \times n}$, called the sensing matrix, with $m \ll n$. We are also given a vector $\mathrm{y} \in \R^m$, which contains a set of $m$ measurements. We are interested in reconstructing $x \in \R^n$, such that $$\mathrm{y} = \Phi x$$ Given that $m \ll n$, this setting is under-determined. However, under the completely reasonable premise that $x$ is compressible, i.e., it is well approximated by $k$-sparse representations, the problem of recovering $x$ becomes feasible. In other words, if we restrict ourselves to vectors which have at most $k$ non zero co-efficients, i.e. $\|x\|_0 = k$ and $k \ll n$, we can efficiently search for solutions. In fact, Candes et al.~\cite{candes2005decoding, candes2006stable, candes2008restricted} show that it is possible recover a $k$-sparse $x$ \emph{exactly} if the sensing matrix $\Phi$ exhibits the \emph{Restricted Isometry Property} (RIP).

\begin{definition}
A matrix $\Phi$ is said to exhibit $(k, \delta)$-RIP iff $\forall x \in \R^n$ with $\|x\|_0 = k$ $$(1-\delta)\|x\|_2 \le \|\Phi x\|_2 \le (1+\delta)\|x\|_2$$
\end{definition}

Please note that $k$ might be referred to as `order' and $\delta$ might be referred to as `Restricted Isometry Constant (RIC)'. Candes et al. showed that it is possible to reconstruct a $k$-sparse $x$ very efficiently, by solving the minimization problem $$\min_{\mathrm{a} \in \R^n} \|\mathrm{a}\|_1 \qquad \qquad \text{subject to }\,\, \Phi \mathrm{a} = \mathrm{y}$$ when we have RIP matrices\footnote{Henceforth, we shall use the phrase `RIP matrices' instead of saying `matrices exhibiting the RIP' every time. It is worth noting that Tao once used the abbreviation `UUP', which stands for `Uniform Uncertainty Principle', for what is now commonly known as RIP.} with $\delta < \sqrt{2} - 1$. Their work has resulted in significant effort towards both deterministic and probabilistic constructions of RIP matrices. Naturally, we want to be able to construct RIP matrices with $k$ as large as possible, but deterministic constructions, such as those presented by Bourgain et al.~\cite{bourgain2011breaking} and DeVore~\cite{devore2007deterministic}, cannot produce RIP matrices of order much greater than $\sqrt{n}$. Deterministic constructions are far from achieving the orders achieved by probabilistic constructions. On the other hand, it has been proven that $\pm 1$ symmetric Bernoulli matrices, or matrices formed by sampling from a Gaussian distribution $\mathcal{N}(0, \frac{1}{n})$, satisfy RIP with $k \in \Theta(n)$\cite{vershynin2010introduction, baraniuk2008simple} with high probability. 

The superiority of random constructions motivates the problem of certifying whether a randomly drawn matrix $\Phi$, from any of the models mentioned above, exhibits RIP with the required parameters. If we find that a randomly drawn matrix is unsuitable to our purposes, we re-generate it and repeat the certification process. Terry Tao posted the following question on his blog~\cite{taoblog}: ``An alternate approach (to deterministic construction of RIP matrix), and one of interest in its own right, is to work on improving the time it takes to verify that a given matrix (possibly one of a special form) obeys the UUP (RIP).'' In this paper, we prove that RIP certification is NP-hard to approximate in a \emph{strong} sense assuming the truth of the \sseh. 

We now state \sseh, which was proposed by Raghavendra and Steurer~\cite{raghavendra2010graph}, and is one of the most important conjectures in complexity theory. In order to present the conjecture, we first define the expansion of a graph.

\begin{definition}Given a graph $d$-regular graph $G(V,E)$ with $n$ vertices, 
we define the expansion of a non-empty set $S\subseteq V$ as $$\phi_G(S) = \frac{|E(S, V - S)|}{d \cdot \mathrm{min}(|S|,\, |V - S|)}.$$
where $E(S,V-S)$ denotes the collection of edges of $G$ that have one end in $S$ and the other end in $V-S$. The expansion of the graph $G$ is defined as the minimum expansion among all subset of its vertices:
        $$\phi(G) = \min_{\substack{S \subseteq V}} \phi(S).$$ 
        
For any $0<\delta \leq 1/2$, we also define the minimum expansion among all subsets of size $\le \delta n$ as $$\phi_\delta(G) = \min_{\substack{S \subseteq V \\ S\leq \delta n}} \phi(S).$$ 
\end{definition}

The Small Set Expansion conjecture states that:

\begin{conj}
\label{conj:sse}
For every $\epsilon > 0$, $\exists\, 0 \le \delta \le \frac{1}{2}$, such that it is NP-hard to distinguish between:
\begin{itemize}
\item $\exists \,S \subseteq V$, with $|S| = \delta|V|$, such that $\phi_G(S) \le \epsilon$
\item  $\forall S \subseteq V$, with $|S| \le \delta|V|$, we have $\phi_G(S) \ge 1 - \epsilon$
\end{itemize}
\end{conj}
\subsection{Our main result}
We give a gap preserving reduction from the \sse problem to the RIP certification problem. More formally, we prove the following theorem:

\begin{theorem}
\label{thm:riphard}
For any $0 \le \delta \le 1$, and $C \ge 1$, there exists $k$ such that, given a matrix $M$ it is \ssehard\footnote{A problem $\mathcal{I}$ is defined to be \ssehard if $\sse \le_{\mathrm{P}} \mathcal{I}$} to distinguish between:

\begin{itemize}
\item (Highly RIP) $M$ is $(k,\delta)$-RIP.
\item (Far away from RIP) $M$ is not $(k/C, 1-\delta)$-RIP.
\end{itemize}

\end{theorem}
We claim that our result has a very strong form, which we will justify in more detail a little later. Also, as corollaries, we have that 

\begin{corollary}Given a matrix $M$ and $k$, it is \ssehard to distinguish whether the matrix is $(k,\delta)$-RIP or not $(k,1-\delta)$-RIP for any $\delta>0$
\end{corollary}

\begin{corollary}Given a fixed $\delta$ and matrix $M$, it is \ssehard to get a constant approximation for the smallest $k$ such that $M$ exhibits $(k,\delta)$-RIP.
\end{corollary}

\subsection{Comparison with Previous work}
Let us go over some previous work on the topic of hardness of RIP certification, and also make a few observations about Theorem~\ref{thm:riphard} to justify our claim that we are proving hardness of RIP certification in a very strong sense. Bandeira et al.~\cite{ieeeriphard} prove that the exact decision version of the problem of RIP certification is $NP$-hard. In other words, they proved that given $\delta, k$, it is NP-hard to certify whether a matrix exhibits $(k,\delta)$-RIP or not. It was later established by Tillmann and Pfetsch~\cite{tillmann2014} that the same problem is also co-NP-hard. Both works reduce from the problem of determining the spark of a matrix, which is known to be NP-hard. It should be mentioned that $\delta$ in both results is in $o_n(1)$, where $n$ is the number of rows of the matrix. Also, we must note that the exact decision version of the problem is restrictive.

Results by Koiran and Zouzias (KZ) are the only works we know of on the approximation version of the problem. KZ obtain various inapproximability results by making assumptions on the hidden clique problem~\cite{KZriphard2} and the dense subgraph problem~\cite{KZriphard1}. Most of the results are of the form that, for some $k,\delta_1,\delta_2$, (depending on the assumption used), it is hard to distinguish whether a matrix is $(k,\delta_1)$-RIP, or not $(k,\delta_2)$-RIP. In almost all of the cases, $\delta_1, \delta_2$ are $\in o_n(1)$, with the exception of one result, which we shall state below:

\begin{itemize}
\item No polynomial time algorithm can distinguish matrices that satisfy the $(\Theta(n), \frac{\kappa}{2})$-RIP from matrices that do not satisfy the $(\Theta(n), \kappa)$-RIP
\end{itemize}

where $\kappa \le \frac{\sqrt{5}}{3}$ is an unknown constant depending on the correctness of hidden-clique and densest-subgraph conjectures. In practice, it is known that an RIP matrix is useful for many applications as long as  $\delta \le \sqrt{2}-1$. Clearly, the above theorem does not rule out the existence of an algorithm for deciding whether the RIC of a matrix is $\le \sqrt{2} - 1$. This is because there is no guarantee that $\kappa \in (\sqrt{2} - 1, 2\sqrt{2} - 2)$. KZ also state ``...our hardness results do not rule out the existence of a polynomial-time algorithm distinguishing between matrices with a very small RIP parameter and matrices with a RIP parameter larger than say 0.1...''. 

Theorem~\ref{thm:riphard} is clearly equipped to make stronger statements than any previous work on inapproximability of RIP parameters. It suggests that certifying RIP for any constant $0<\delta<1$ is \ssehard. In addition, even if the matrix indeed exhibits \emph{strong} RIP (small constant $\delta$ and very large $k$), it is still \ssehard to even certify if it exhibits \emph{weak RIP} (with large $\delta$ close $1$ and small $k$).

\subsection{Proof Overview}
Let us assume that $G$ is a $d$-regular graph with adjacency matrix $A$, and $L = I -\frac{1}{d}A$ is the normalized Laplacian matrix of the graph. it is easy to see that, given $x_S\in \{0,1\}^n$ as the indicator vector of set $S$, we have that 
$$
        \phi(S) =\frac{x_S^T L x_S}{\xs\trans\xs}= \frac{\|Mx_S\|_2^2}{\|x_S\|_2^2}
$$
for $M$ satisfying $M\trans M = L$. We know that the Laplacian is a quadratic form and thus is positive semi-definite. Thus, $L$ always admits the decomposition $L = M\trans M$.

The strategy of the reduction is to take a \sse instance and to construct the corresponding $M$ for the RIP certification problem. Our reduction has a similar flavor to the reduction in Koiran and Zouzias~\cite{KZriphard2} (they call their reduction as \emph{Cholesky Reduction}). If there is a small set $S$ with expansion less than $\eps$, we know that $x_S$, corresponding to this $S$, is a sparse vector such that $\|Mx_S\|_2 \leq \sqrt{\eps}\|x_S\|_2$ and this suggests that $M$ is far from being a RIP matrix. The second case of the proof is more involved. Here, we would like to show that if there exists a $k$-sparse $x\in \R^n$ such that $x\trans L x$ is bounded away from $1$, then we can find a small set $S$ such that $\phi(S)$ is also bounded away from $1$. 

If we ignore the sparsity constraint, this kind of conversion from a real vector $x$ to a boolean vector $x_S$ is exactly reminiscent of the ``hard direction'' of Cheeger's Inequality~\cite{polya1951isoperimetric, sinclair1989approximate, alon1986eigenvalues}. In this paper, we prove the following generalization of Cheeger's Inequality for sparse vectors, which we use to prove the ``hard direction''. 

\begin{theorem} 
\label{thm:sparse}
(Sparse Cheeger's Inequality) Let $A$ be the adjacency matrix of a $d$-regular graph $G$, and $L= I - \frac{1}{d}A$ be its normalized Laplacian matrix. For any $\delta \leq 1/2$, we have that $${\lambda_{\delta}} \le \phi_{\delta}(G)\leq   \sqrt{(2-\lambda_\delta)\lambda_\delta}$$ 
where $\lambda_{\delta} = \min_{|x|_0=\delta |V|} \frac{x\trans  L x}{x\trans x}$
\end{theorem}

The above inequality establishes a relationship between the minimum expansion of $G$ on small sets with the minimum value of $\frac{x\trans L x}{x\trans x}$ for sparse real vectors $x$. A similar and independent, but not identical, result is known - Theorem 2.1 in Steurer~\cite{steurer2010subexponential}. We observe that the constants in Theorem~\ref{thm:sparse} are better, and this might find applications elsewhere.

For the purpose of comparison, we also list the original Cheeger's Inequality here:

\begin{theorem} 
\label{thm:cheeger}
(Cheeger's Inequality) Let $A$ be the adjacency matrix of a graph $G$, and $L= I - \frac{1}{d}A$ be its normalized Laplacian matrix. We have that $$\frac{\lambda}{2} \le \phi(G)\leq   \sqrt{2\lambda}$$ 
where $$\lambda = \min_{\substack{x \in \R^n \\ Lx \ne \vec{0}}} \frac{\|x^T L x\|_2}{\|x\|_2^2}$$ is the second smallest eigenvalue of $L$.
\end{theorem}

 It is interesting to note here that the relationship between $\lambda_{\delta}$ and $\phi_{\delta}(G)$ in Theorem~\ref{thm:sparse} is tighter than the relationship between $\lambda$ and $\phi(G)$ in Theorem~\ref{thm:cheeger}.  It is crucial for our proof that we get $\sqrt{(2-\lambda_\delta)\lambda_\delta}$ instead of $\sqrt{2\lambda_\delta}$ in Theorem~\ref{thm:sparse}. We need to prove that if there exists a $\delta$-sparse vector $x$\footnote{$\|x\|_0 \le \delta n$} such that  $\lambda_{\delta} $ is bounded away from $1$, then there is a small set whose expansion is also bounded away from 1. If what we had was only $\phi_{\delta}(G)\leq \sqrt{2\lambda_\delta}$, the bound becomes trivial even when we know that $\lambda_{\delta}=1/2$. It is only because of Theorem~\ref{thm:sparse} that we find that, as long as  $\lambda_{\delta}$ is bounded away from $1$, we also have that $\sqrt{(2-\lambda_\delta)\lambda_\delta}$ is bounded away from $1$. This turns out to be exactly what we need to prove.

The proof of the sparse Cheeger's Inequality bears resemblance to the proof of the classical Cheeger's Inequality (e.g., see~\cite{luc}). The proof makes it necessary to strengthen the analysis for the sparse vector case so as to obtain a tighter relationship between $\lambda_{\delta}$ and $\Phi_{\delta}(G)$.

One final thing to notice is that our hardness result amplifies the dependence on the order $k$. We show that it is hard to distinguish $(k,1-\delta)$-RIP from $(k/C,\delta)$-RIP, where $C$ is any arbitrary constant. To this end, we need to use an equivalent statement of the \sseh by Raghavendra et al.~\cite{RST12} which gives us a stronger starting point for the hardness reduction.

\subsection{Organization}
The paper is organized as follows. In Section~\ref{sec:cheeger}, we prove the sparse Cheeger's Inequality, namely Theorem~\ref{thm:sparse}. We then use this theorem to prove our main result, i.e Theorem~\ref{thm:riphard} which is presented in Section~\ref{sec:hardness}.

%
%

\section{Sparse Cheeger's Inequality}
\label{sec:cheeger}

Below we state the proof of Theorem~\ref{thm:sparse}.

\begin{proof}
Assuming that $|V|=n$,  let us first prove the left side of the inequality, analogous to what is commonly called the \emph{easy direction}. Choose $x_S \in \{0,1\}^n$, as a bit vector representation of a set $S$ of size at most $\delta |V|$.  We then easily get
$$
\phi_{\delta}(G) = \frac{x_S^T L x_S}{\|\xs\|^2} \leq \lambda_{\delta}
$$

Next, we prove the right hand side of inequality in Theorem~\ref{thm:sparse}. Assume we are given any $x \in \R^n$ such that $$\frac{x^TLx}{\|x\|^2_2}=\lambda_\delta$$ We shall prove that, using $x$, we will be able to construct some set $S$ such that $\phi(S)\leq \sqrt{(2-\lambda_{\delta})\lambda_\delta}$. This will complete the proof because $\phi_{\delta}(G)$ is the minimum value of $\phi(S)$ over all $S$ with $|S| \le \delta n$. \\

Let us use $x_i$ to indicate the $i$-th coordinate of $x$, by the property of the Laplacian matrix of a graph, we know that 
\begin{equation}
\frac{x^TLx}{\|x\|_2^2 } = \frac{\sum_{1\leq i,j\leq n} (x_i-x_j)^2 A_{ij}}{2d\sum_{i=1}^n |x_i|^2} \label{eqn:lap}
\end{equation}

Without loss of generality, we can assume that all coordinates $x_i$ have nonnegative value because changing $x_i$ to $|x_i|$ does not increase $\sum_{1\leq i,j\leq n} (x_i-x_j)^2 A_{ij}$ and $\sum |x_i|^2$ remains the same. Also, we can scale $x$ such that $0\leq x_1\leq x_2 \ldots\leq x_n=1$ because $\frac{x^TLx}{\|x\|^2_2}$ does not change when we scale $x$.

Consider a distribution $P$ with density  $f(x)= 2x$ on the interval $(0,1)$. It is easy to verify this is a valid density function as
$$
        \int_{0}^1 2x \cdot \mathrm{d}x=1.
$$
  Now consider the following randomized construction of set $S$ from $x$.

\begin{enumerate}
\item Choose $t$ in $(0, 1)$ according to $P$
\item Set $S$ to be $S_t=\{ i \ |\ x_i  \geq t\}$
\end{enumerate}

Given the fact that $x$ is $\delta n$ sparse, $|S_t|\leq \delta n$ for any $0<t<1$. We can easily see that for any $0 \le a \le b \le 1$, we have $$\Pr\left(t\in [a,b]\right) =\int_{a}^{b}2x dx =  b^2 -a^2.$$
Therefore, $$\Pr(x_i \in S_t)= \Pr(x_i\geq t)=x_i^2$$  
which implies that   
$$\E_{t}\left[|S_t|\right] = \sum_{i=1}^n x_i^2$$ 

Also we know that for any $i,j$, there is an edge between vertex $i$ and vertex $j$ only if $t$ falls between $x_i$ and $x_j$.
Therefore, 
\begin{multline*}\E_t\left[|E(S_t, V - S_t)|\right]= \frac{1}{2}\sum_{i,j} A_{ij} |x_i^2 - x_j^2|=\frac{1}{2}\sum_{1\leq i,j\leq n} A_{ij} |x_i-x_j||x_i + x_j|\\
\leq \sqrt{\frac{\sum_{1\leq i,j\leq n} A_{ij}(x_i+x_j)^2}{2}}\cdot \sqrt{\frac{\sum_{1\leq i,j\leq n} A_{ij}(x_i-x_j)^2}{2}} $$
\end{multline*}
The last inequality in the above sequence of steps is due to the Cauchy-Schwarz Inequality. We then calculate ratio between $\E_t\left[|E(S_t, V - S_t)|\right]\text{ and }d \cdot \E_t\left[|S_t|\right]$, which is $$\frac{\E_t\left[|E(S_t, V - S_t)|\right]}{d \cdot \E\left[|S_t|\right]}\leq \sqrt{\frac{\sum_{1\leq i,j\leq n} A_{ij}(x_i+x_j)^2}{2d\sum_{i=1}^n x_i^2}}\sqrt{\frac{\sum_{1\leq i,j\leq n} A_{ij}(x_i-x_j)^2}{2d\sum_{i=1}^n x_i^2}}.
$$

By Equation~\eqref{eqn:lap}, we know that $$ \frac{\sum_{1\leq i,j\leq n} A_{ij}(x_i-x_j)^2}{2d\sum_{i=1}^n x_i^2} = \frac{x^T L x}{|x|^2 } = \lambda_{\delta} $$     

We also know that $$ \frac{ \sum_{1\leq i,j\leq n} A_{ij}(x_i+x_j)^2}{2d\sum_{i=1}^n x_i^2} + \frac{ \sum_{1\leq i,j\leq n} A_{ij}(x_i-x_j)^2}{2d\sum_{i=1}^n x_i^2} = \frac{ \sum_{1\leq i,j\leq n} A_{ij}(2x_i^2+2x_j^2)}{2d\sum_{i=1}^n x_i^2}=2 $$ which implies that $$\frac{ \sum_{1\leq i,j\leq n} A_{ij}(x_i+x_j)^2}{2d\sum_{i=1}^n x_i^2} = 2-\lambda_{\delta}$$
 
This suggests that $$ \frac{\E_t[|E(S_t, V - S_t)|]} {d \cdot \E_t[|S_t|]}\leq \sqrt{\lambda_{\delta}(2-\lambda_\delta) } $$ or equivalently $$\E_t\left[|E(S_t, V - S_t)|-\sqrt{\lambda_{\delta}(2-\lambda_\delta) } \cdot d \cdot |S_t|\right]\leq 0 $$

Therefore, there must exist some $t\in (0,1)$ such that $$ |E(S_t, V - S_t)|-\sqrt{\lambda_{\delta}(2-\lambda_\delta) } \cdot d \cdot |S_t|\leq 0 $$ or in other words $$\phi(S_t) = \frac{|E(S, V - S)|}{d \cdot |S_t|}\leq \sqrt{\lambda_{\delta}(2-\lambda_\delta) }$$

Therefore, if we choose the best $t\in (0,1)$, we know that $$\min_t \phi(S_t)\leq \sqrt{\lambda_{\delta}(2-\lambda_\delta) }$$

This finishes the proof for the right hand side of the inequality in Theorem~\ref{thm:sparse}.
\end{proof}

\section{Proof for the Hardness of Certifying RIP}
\label{sec:hardness}

\subsection{Equivalent variant of the \sseh}~\label{sec:sse}
The starting point is the following Theorem~\ref{thm:strongsse} from~\cite{RST12}, which states that a strengthened form of \sseh is equivalent to the original \sseh.

\begin{theorem}
\label{thm:strongsse}
Given a $d$-regular graph $G(V,E)$, for all constant integer $q\in \N$, and any constant $\gamma,\eps\in (0,1)$, it is \ssehard to distinguish the following two cases:

\begin{itemize}
\item there are q disjoint sets $S_1,S_2, \ldots, S_q\subseteq V$ of size $\frac{n}{q}$ such that $\phi(S)\leq \epsilon+o(\epsilon)$.
\item for any $0<\beta<1$, every set of $S\leq \beta n, S\subseteq V$  has expansion at least $1-\frac{T_{1-\eps}(\beta)}{\beta}-\gamma/\beta$
\end{itemize}
\end{theorem}

Here $T_{1-\eps}$ is related to the Gaussian Stability function, which is defined by Khot et al. in~\cite{KKMO07}. We will use the following upper bound that was presented in~\cite{KKMO07}:

\begin{equation}
\frac{T_{1-\eps/2}(\beta)}{\beta} \leq \beta^{\eps/4} \label{eq:tailbd}
\end{equation}

for any $\beta,\eps$. By putting $\beta= {(\eps)^{4/\eps}},\alpha = \beta/C, q=1/\alpha,  \gamma =  \eps^{4/\eps+2}$ in inequality~\eqref{eq:tailbd}, we have that 
$$\frac{T_{1-\eps}(\beta)}{\beta}+\gamma/\beta \leq \eps+ o(\eps)$$
and the following hardness statement of \sse:
\begin{theorem}
\label{thm:start} 
For any $0<\eps<1$, and an arbitrarily large constant $C$, there exists some $k<n$ (functionally dependent on $\eps$), for which it is \ssehard to distinguish the following two cases in a $d$-regular graph $G(V,E)$:

\begin{itemize}
\item there is a set $S\subseteq V$ of size $k/C $ such that $\phi(S)\leq O(\eps)$
\item every set $S\subseteq V$ of size  less than $k$  has expansion at least $ 1-O(\eps)$
\end{itemize}
\end{theorem}
\subsection{Hardness Reduction}\label{sec:reduction}
We shall make a gap preserving reduction from the \sse hardness of Theorem~\ref{thm:start}. Given any $d$-regular graph with adjacency matrix $A$, will consider matrix $M$ such that $M\trans M = I - \frac{1}{d}A$ for the RIP certification problem. Also without loss of generality, we can only prove for  $\delta$ that is sufficiently small constant as if Theorem~\ref{thm:riphard} holds for  some $\delta=\delta_0$, it also holds for all $\delta\geq \delta_0$.
In order to prove Theorem~\ref{thm:riphard}, it suffices to prove the following Lemma~\ref{lem:reduction}.
 
\begin{lemma}
~\label{lem:reduction}  
Let $\delta =\eps^{0.4}$ for a sufficiently small constant $\eps$. Then:
\begin{enumerate}
 \item If there is a set $S$ of size at most $k/C$ and $\phi_G(S)\leq O(\eps)$, then the matrix is $M$ not $(k/C,1-\delta)$-RIP
\item If for every set S of size at most $k$, $\phi_G(S)\geq  1-O(\eps)$, then $M$ is $(k,\delta)$-RIP
\end{enumerate}
\end{lemma}

The proof of Lemma~\ref{lem:reduction} would complete the proof of Theorem~\ref{thm:riphard}.

\begin{proof}
Given any $d$-regular graph with adjacency matrix $A$, let $x_S\in \{0,1\}^n$ be the indicator vector of a subset $S$. We know the number of edges that leave $S$ is equal to $d\cdot |S|- x_S\trans A x_S = x_S\trans(d\cdot I-A )x_S$.  Therefore, we have
\[
        \phi_G(S) = \frac{x_S\trans(d\cdot I- A ) x_S}{d |S| }  =  \frac{x_S\trans(d\cdot I- A ) x_S}{d \|x_S\|_2^2 } =\frac{x_S\trans( I- A/d) x_S}{ \|x_S\|_2^2 }=\frac{x_S\trans M\trans M x_S}{ \|x_S\|_2^2 }
\]
Let us prove the first claim. We know that when there is a set $S\subseteq V$ of size less than $\frac{k}{C}$ that has expansion less than $O(\epsilon)$. Let us denote $x_S\in \{0,1\}^n$ as the indicator of set $S$, then $$\frac{x_S\trans \cdot M\trans M (x_S)}{\|x_S\|^2} \leq O(\eps)$$ 
which implies that 
$$\|M x\|_2 \leq O(\sqrt{\eps})\|x\|_2 $$
  
 Since $x_S$ is $k/C$-sparse, after applying $M$, its length is only $O(\sqrt{\eps})$ times $\|x\|_2$. Now, given that we know $\delta = \eps^{0.4}$, for sufficiently small $\eps$, we have that $M$ is not $(k/C,\delta)$-RIP.

We shall prove the second claim of Lemma~\ref{lem:reduction} by contradiction.  Suppose there exists some $k$-sparse vector $x\in \R^n$ such that $$\|Mx\|_2 \leq (1-\delta) \|x\|_2$$ We know then that $$x\trans M\trans M x \leq (1-\delta)^2 \|x\|_2,$$which implies that $\lambda_\delta(G)\leq  1-2\delta +\delta^2$. Now, by Theorem~\ref{thm:sparse}, we have that there must exist a set such that the expansion is at most $$\sqrt{\lambda_{\delta}(G)(2-\lambda_\delta(G))}\leq \sqrt{1-(2\delta -\delta^2)^2} = 1-\Theta(\eps^{0.8}),$$
which contradicts the fact that all sets $S$ of size less than $k$ must have expansion at least $1-\eps$.
\end{proof}

\section{Conclusion and Open Problems}

In this paper, we establish that certifying RIP of a matrix (even approximately) is \ssehard in a strong sense. Although the \sse problem is a conjecture, our work helps cement the place of RIP certification relative to other problems in regard to their hardness. In general, whenever we reduce from a known problem to a new problem, it increases the importance of the original problem.

One possible immediate open problems is to prove NP-hardness of RIP certification by reducing from known canonical problems. This would be interesting and important because the correctness of \sseh is uncertain. Another interesting direction to pursue could be to prove that RIP certification is hard even when the matrix satisfies certain natural properties such as coherence.

\subsection*{Acknowledgements}

The authors would like to thank the anonymous reviewers for their constructive and insightful comments which helped in the final presentation of the paper.
\bibliographystyle{plain}
\bibliography{main}

\end{document}